 \documentclass[final,5p,times,twocolumn]{elsarticle}

\usepackage{amsmath}
\usepackage{amssymb}
\usepackage{booktabs}
\usepackage{color}
\usepackage{graphicx}
\usepackage{hyperref}
\usepackage{subfig}
\usepackage{xspace}

\newcommand{\pygbe}{\texttt{PyGBe}\xspace}
\newcommand{\cpu}{\textsc{cpu}}
\newcommand{\gpu}{\textsc{gpu}}
\newcommand{\bem}{\textsc{bem}\xspace}

\newcommand{\nvidia}{\textsc{nvidia}\xspace}
\newcommand{\msms}{\texttt{\textsc{msms}}\xspace}
\newcommand{\apbs}{\textsc{\texttt{apbs}}\xspace}
\newcommand{\afmpb}{\textsc{\texttt{afmpb}}\xspace}
\newcommand{\mibpb}{\textsc{\texttt{mibpb}}\xspace}
\newcommand{\tabi}{\textsc{\texttt{tabi}}\xspace}
\newcommand{\gmres}{\texttt{\textsc{gmres}}\xspace}
\newcommand{\mac}{\textsc{mac}\xspace}
\newcommand{\ccby}{\textsc{cc-by}\xspace}

\newcommand\ignore[1]{} 
\newcommand{\comment}[1]{\textcolor{red}{ #1 }}
\ignore{
\renewcommand{\comment}[1]{}
}

\journal{Computer Physics Communications}

\begin{document}

\begin{frontmatter}



\title{A biomolecular electrostatics solver using Python, GPUs and boundary elements that can handle solvent-filled cavities and Stern layers}

\author[bu]{Christopher D. Cooper}
\ead{cdcooper@bu.edu}

\author[neu]{Jaydeep P. Bardhan}
\ead{jbardhan@ece.neu.edu}

\author[bu]{L.~A.~Barba\corref{lab}}
\ead{labarba@gwu.edu}

\address[bu]{Mechanical Engineering, Boston University, Boston, MA, 02215 U.S.A.}
\address[neu]{Electrical and Computer Engineering, Northeastern University, Boston, MA, 02115 U.S.A.}
\cortext[lab]{New address:  Mechanical and Aerospace Engineering, George Washington University, Washington, D.C., 20052,
\href{mailto:labarba@gwu.edu}{labarba@gwu.edu}}

\begin{abstract}

The continuum theory applied to biomolecular electrostatics leads to an implicit-solvent model governed by the Poisson-Boltzmann equation. Solvers relying on a boundary integral representation typically do not consider features like solvent-filled cavities or ion-exclusion (Stern) layers, due to the added difficulty of treating multiple boundary surfaces. This has hindered meaningful comparisons with volume-based methods, and the effects on accuracy of including these features has remained unknown. This work presents a solver called \pygbe that uses a boundary-element formulation and can handle multiple interacting surfaces. It was used to study the effects of solvent-filled cavities and Stern layers on the accuracy of calculating solvation energy and binding energy of proteins, using the well-known \apbs finite-difference code for comparison. The results suggest that if required accuracy for an application allows errors larger than about 2\% in solvation energy, then the simpler, single-surface model can be used. When calculating binding energies, the need for a multi-surface model is problem-dependent, becoming more critical when ligand and receptor are of comparable size. Comparing with the \apbs solver, the boundary-element solver is faster when the accuracy requirements are higher. The cross-over point for the \pygbe code is in the order of 1--2\% error, when running on one \gpu\ card (\nvidia Tesla C2075), compared with \apbs running on six Intel Xeon \cpu\ cores. \pygbe achieves algorithmic acceleration of the boundary element method using a treecode, and hardware acceleration using \gpu s via PyCuda from a user-visible code that is all Python. The code is open-source under MIT license.

\end{abstract}


\end{frontmatter}


\section{Introduction}
\label{sec:intro}

Many vital biomolecular processes are dominated by electrostatic forces, such as solvation and binding. These processes occur in an aqueous environment, and it is crucial to include the effect of water (the solvent) in any model. In biology, salt ions in the solvent always play a role in the electrostatic potential field, and they must also be considered in the model. Thus, simulation of biomolecular systems is a challenging task, in particular if one aims for a detailed description of all molecules in the system, such as in molecular dynamics simulations. In some situations, the systems of interest are too large for a detailed description using molecular dynamics; in that case, we can look at mean-field potentials instead, using an implicit-solvent model \cite{Roux1999,Bardhan2012}.

Implicit-solvent models rely on the solution of the Poisson-Boltzmann equation. Research through the years on the numerical solution of this equation has led to several codes that are open and widely used in the community, such as \apbs \cite{BakerETal2001}, Delphi \cite{GilsonSharpHonig1987}, \textsc{\texttt{mead}} \cite{BashfordGerwert1992,Bashford1997}, \mibpb \cite{GengYuWei2007}, \afmpb \cite{LuETal2006}, \tabi \cite{GengKrasny2013}, among others. The most widely used are volumetric-based solvers, meaning that they create a discretization of the volume in which the equations apply. On the other hand, boundary-element methods (\bem), which discretize only surfaces, have become a popular alternative in biomolecular applications \cite{YoonLenhoff1990,JufferETal1991,Zhou1993} and several \bem codes have been developed in recent years \cite{LuETal2006,GengKrasny2013,AltmanBardhanWhiteTidor05,BajajETal2011}. 

In \bem solvers, it is more difficult to consider features like solvent-filled cavities and Stern layers around biomolecules, compared with volumetric-based solvers. In fact, it appears that apart from the work of Altman and co-workers \cite{AltmanBardhanWhiteTidor09}, no other \bem implementation addresses solvent-filled cavities and Stern layers. Consequently, the literature does not explain the importance of including these features in a model. Moreover, without boundary-integral formulations that treat these features, comparisons between \bem and volumetric-based methods are not meaningful, because the latter always include them.

In this paper, we study the effect of including solvent-filled cavities and Stern layers in a Poisson-Boltzmann \bem solver. This allows us to make a fair comparison with a volumetric-based method, in this case \apbs, in order to determine in which cases a boundary integral approach may be preferable.
To perform the study, we developed \pygbe, a treecode-accelerated boundary element solver \cite{CooperBarba-share154331}. \pygbe is written in Python and uses PyCUDA \cite{Kloeckner2012} to exploit \gpu\ hardware for shorter runtimes. The code is free and open-source under an MIT license, and we welcome interactions with the community. 

\vspace{1cm}

\section{Methods}
\label{sec:method}

\subsection{Implicit-solvent model} \label{sec:model}

The implicit-solvent model describes the electrostatic field in a molecular system using continuum theory. We define a solvent region (water with salt) and a protein region, separated by the solvent-excluded surface (SES)~\cite{Connolly1983a}. The solvent region has a high dielectric constant (permittivity $\sim80$) and is modeled by the linearized Poisson-Boltzmann equation to account for the presence of ions. The protein region has low permittivity ($\sim2-4$) and contains point charges at the locations of the atoms, and it is modeled by the Poisson equation. 

\begin{figure}[t]
   \centering
   \includegraphics[width=0.64\columnwidth]{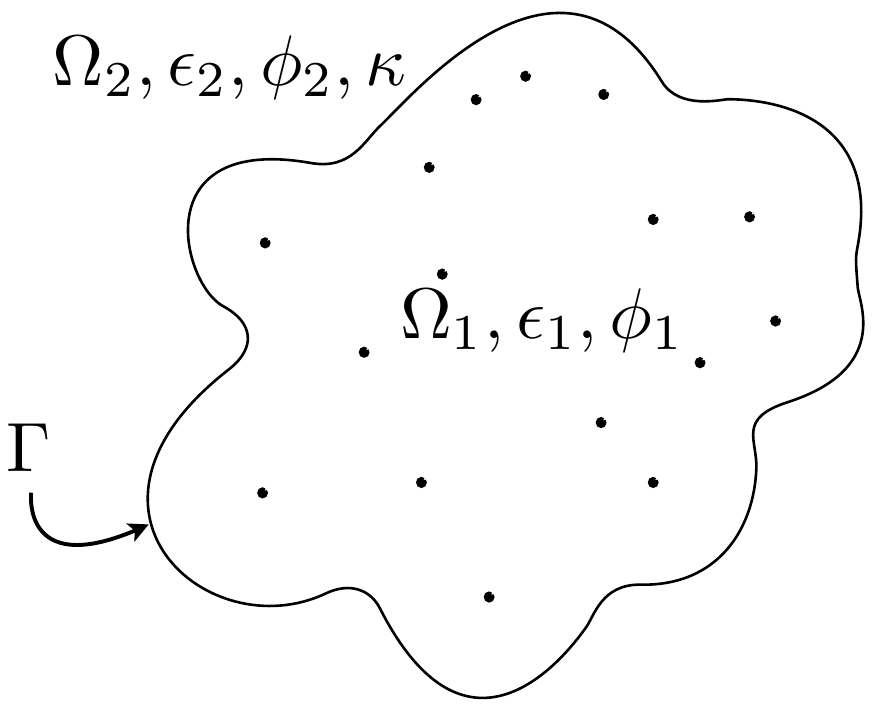}
   \caption{Regions in the implicit-solvent model: $\Omega_1$ is the protein region, with permittivity $\epsilon_1$ and potential $\phi_1$; $\Omega_2$ is the solvent region, with permittivity $\epsilon_2$, inverse Debye length $\kappa$, and potential $\phi_2$; $\Gamma$ is the solvent-excluded surface and the dots represent the locations of atoms.}
   \label{fig:molecule_model}
\end{figure}

\begin{figure}
   \centering
   \includegraphics[width=0.64\columnwidth]{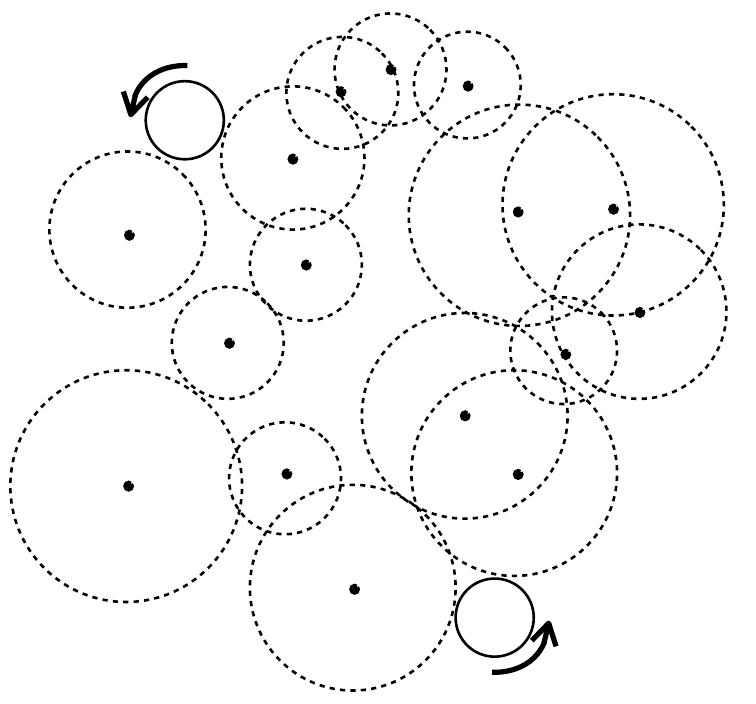} 
   \caption{Generation of the solvent-excluded surface. The dots are the locations of the atoms and the dotted circles represent the atomic spheres with appropriate radii. The solid ball represents the spherical probe used to compute the SES.}
   \label{fig:molecule_atom}
\end{figure}

Figure \ref{fig:molecule_model} depicts the two regions. The SES represents the closest a water molecule can get to the protein.
It is defined by rolling a probe sphere of the size of a water molecule ($1.4$~\AA) around the protein, described by the atoms with their corresponding van der Waals radius.  This process is sketched in Figure \ref{fig:molecule_atom}. 

Using the notation from Figure \ref{fig:molecule_model}, the electrostatic potential can be described by the following system of partial differential equations:

\begin{align} \label{eq:pde}
\nabla^2 \phi_1(\mathbf{r}) &= - \sum_k \frac{q_k}{\epsilon_1} \delta(\mathbf{r},\mathbf{r}_k) \ \text{ in protein region $\Omega_1$}  \nonumber \\ 
\nabla^2\phi_2 (\mathbf{r}) &= \kappa^2 \phi_2(\mathbf{r}) \quad \qquad \ \ \text{ in solvent region $\Omega_2$}  \nonumber \\ 
\phi_1 &=\phi_2 \qquad \qquad \ \ \quad \text{ on interface $\Gamma$}  \nonumber \\ 
\epsilon_1 \frac{\partial \phi_1}{\partial \mathbf{n}} &= \epsilon_2 \frac{\partial \phi_2}{\partial \mathbf{n}} 
\end{align}

\noindent where $q_k$ is the charge associated to atom $k$ and $\kappa$ is the inverse of the Debye length. The boundary conditions in \eqref{eq:pde} enforce continuity of the potential and normal electric displacement through the interface $\Gamma$.

\paragraph{Integral formulation} \label{sec:model_integral}

 If we take the convolution of the PDEs in \eqref{eq:pde} with their corresponding free-space Green's functions, then evaluate the expressions (via integration by parts) on the interface $\Gamma$ applying the continuity conditions, the system is transformed into a system of integral equations along $\Gamma$, as shown by Yoon and Lenhoff \cite{YoonLenhoff1990}:

\begin{align} \label{eq:nBEM}
\frac{\phi_1(\mathbf{r})}{2} - &\oint_{\Gamma} \frac{\partial \phi_1}{\partial \mathbf{n}} (\mathbf{r}') \frac{1}{4\pi|\mathbf{r} - \mathbf{r}'|} \mathrm{d} \Gamma'  + \nonumber \\
	&\oint_{\Gamma} \frac{\partial}{\partial \mathbf{n}} \left[ \frac{1}{4\pi|\mathbf{r} - \mathbf{r}'|} \right] \phi_1(\mathbf{r}') \mathrm{d} \Gamma'  = \sum_{k=0}^{N_c} \frac{q_k}{\epsilon_1} \frac{1}{4\pi|\mathbf{r} - \mathbf{r}_k|}, \nonumber \\
\frac{\phi_1(\mathbf{r})}{2} + \frac{\epsilon_2}{\epsilon_1} &\oint_{\Gamma} \frac{\partial \phi_1}{\partial \mathbf{n}} (\mathbf{r}') \frac{e^{-\kappa |\mathbf{r}-\mathbf{r}'|}}{4\pi|\mathbf{r} - \mathbf{r}'|} \mathrm{d} \Gamma'  - \nonumber \\
	&\oint_{\Gamma} \frac{\partial}{\partial \mathbf{n}} \left[ \frac{e^{-\kappa |\mathbf{r}-\mathbf{r}'|}}{4\pi|\mathbf{r} - \mathbf{r}'|} \right] \phi_1(\mathbf{r}') \mathrm{d} \Gamma' = 0.
\end{align}

\noindent Here, $\frac{1}{4\pi|\mathbf{r}-\mathbf{r}'|}$ and $\frac{e^{-\kappa|\mathbf{r}-\mathbf{r}'|}}{4\pi|\mathbf{r}-\mathbf{r}'}$ are the free-space Green's functions of the Laplace and linearized Poisson-Boltzmann equations, respectively. The position vectors $\mathbf{r}$ and $\mathbf{r}'$ are defined along the boundary. The singularity is treated via a limiting process  in the classic way \cite{CourantHilbert-Vol1}.

\paragraph{Solvation energy} \label{sec:solv_energy}

The electrostatic contribution to the free energy of solvation---the solvation energy---is a common parameter of interest that can be computed from the electric potential. It is defined as the energy required to move the protein from a reference state, often a uniform dielectric, i.e., $\epsilon_1$ in the exterior region, to the solvated state. The total electrostatic potential is the sum of the direct Coulomb potential and the \textit{reaction potential}, which arises due to the presence of the two dielectric materials:
\begin{equation} \label{eq:reac}
\phi_1(\mathbf{r}) =  \phi_{\text{Coul}}(\mathbf{r}) + \phi_{\text{reac}}(\mathbf{r}).
\end{equation}

Since $\phi_{\text{Coul}}$ is exactly the right hand side of the top equation in \eqref{eq:nBEM}, we can compute $\phi_{\text{reac}}$ anywhere in $\Omega_1$ using the following integral relation:
\begin{equation}\label{eq:reac_integral}
\phi_{\text{reac}}(\mathbf{r}) = \oint_{\Gamma} \frac{\partial \phi_1}{\partial \mathbf{n}}(\mathbf{r}') \frac{1}{4\pi|\mathbf{r}-\mathbf{r}'|}  \mathrm{d} \Gamma' - \oint_{\Gamma} \frac{\partial }{\partial \mathbf{n}} \left[\frac{1}{4\pi|\mathbf{r}-\mathbf{r}'|} \right] \phi_1(\mathbf{r}') \mathrm{d} \Gamma'.
\end{equation}

We obtain \eqref{eq:reac_integral} using the same derivation described for Equation \eqref{eq:nBEM}. However, in this case, $\mathbf{r}$ is not taken to the interface $\Gamma$, but is valid anywhere in the domain $\Omega_1$, and $\phi_{Coul}$ is subtracted out.

The energy due to any potential is:
\begin{equation} \label{eq:energy}
\Delta G = \int_{\Omega} \rho \phi d\Omega,
\end{equation}

\noindent where $\rho$ is the charge distribution. In this case, solvation energy is generated by $\phi_{reac}$ and the integral can be computed analytically since $\rho$ is a collection of point charges. Equation \eqref{eq:energy} becomes:
\begin{equation} \label{eq:solv}
\Delta G_{\text{solv}} = \frac{1}{2} \sum_{k=1}^{N_c} q_k \phi_{\text{reac}}(\mathbf{r}_k).
\end{equation}

\paragraph{Binding energy} \label{sec:bind_energy}

We can calculate the electrostatic contribution to the binding free energy ---referred to as binding energy--- with the following relation:
\begin{equation} \label{eq:bind_energy}
\Delta \Delta G_{\text{bind}} = \Delta G_{\text{complex}} - \Delta G_{\text{receptor}} - \Delta G_{\text{ligand}} ,
\end{equation}

\noindent where $\Delta G_i$ is the total electrostatic energy of species $i$:
\begin{equation} \label{eq:total_energy}
\Delta G_i = \Delta G_{i,\text{solvation}} + \Delta G_{i,\text{Coulomb}}.
\end{equation}

\noindent where $\Delta G_{\text{solvation}}$ is calculated as described above and $\Delta G_{\text{Coulomb}}$ corresponds to the Coulombic energy of the point charges.

We approximate the binding energy of a complex assuming that the geometry of the receptor and the ligand remain constant from the unbound to the bound states. 

\subsection{Boundary element method (BEM)} \label{sec:bem}

\paragraph{Single-surface problems} \label{sec:bem_single}

\begin{figure}
   \centering
   \includegraphics[width=0.8\columnwidth]{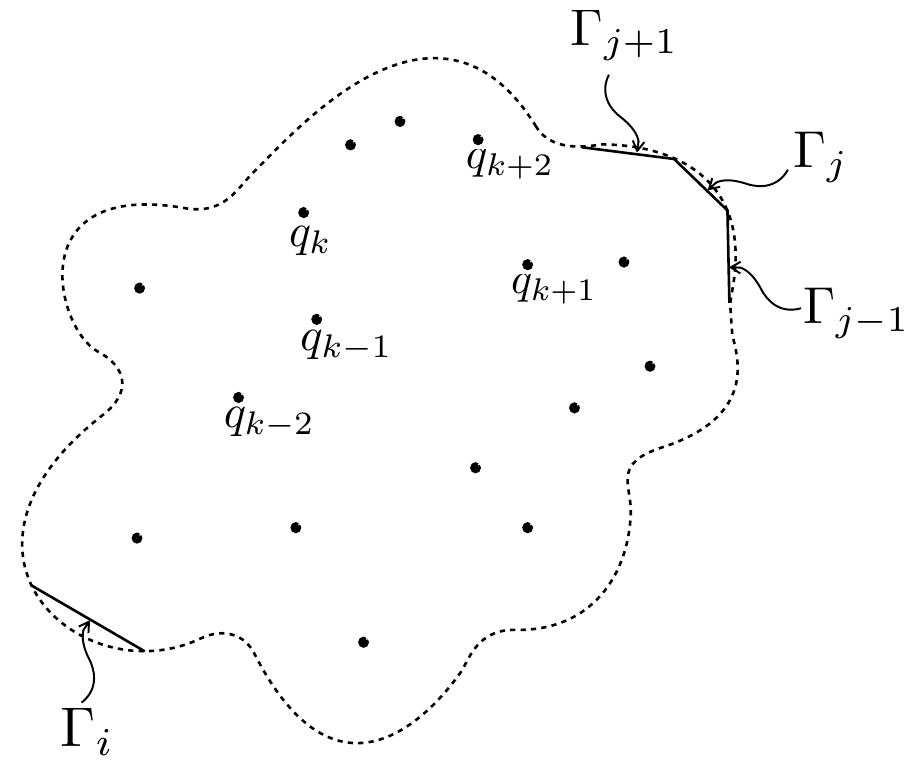} 
   \caption{Sketch of the discretized interface $\Gamma$: $\Gamma_i$ is the panel containing the collocation point and $\Gamma_j$ the panels over which we integrate.}
   \label{fig:molecule_disc}
\end{figure}

In BEM, we discretize the boundary into panels as sketched in Figure \ref{fig:molecule_disc},and assume a distribution of the unknown quantities within those panels. We use flat triangular panels and take $\phi_1$ and $\frac{\partial \phi_1}{\partial \mathbf{n}}$ to be constant on each one of them. Then, we use central collocation to obtain a linear system. The discretized form of \eqref{eq:nBEM} becomes

\begin{align} \label{eq:nBEM_disc}
\frac{\phi_1(\mathbf{r}_i)}{2} - &\sum_{j=1}^{N_p} \frac{\partial \phi_1}{\partial \mathbf{n}} (\mathbf{r}_j) \int_{\Gamma_j}  \frac{1}{4\pi|\mathbf{r}_i - \mathbf{r}'|} \mathrm{d} \Gamma'  +  \nonumber \\
&\sum_{j=1}^{N_p} \phi_1(\mathbf{r}_j)  \int_{\Gamma_j} \frac{\partial}{\partial \mathbf{n}} \left[ \frac{1}{4\pi|\mathbf{r}_i - \mathbf{r}'|} \right] \mathrm{d} \Gamma'  = \sum_{k=0}^{N_c} \frac{q_k}{\epsilon_1} \frac{1}{4\pi|\mathbf{r}_i - \mathbf{r}_k|} \nonumber \\
\frac{\phi_1(\mathbf{r}_i)}{2} + &\sum_{j=1}^{N_p} \frac{\partial \phi_1}{\partial \mathbf{n}} (\mathbf{r}_j)  \frac{\epsilon_1}{\epsilon_2} \int_{\Gamma_j} \frac{e^{-\kappa |\mathbf{r}_i-\mathbf{r}'|}}{4\pi|\mathbf{r}_i - \mathbf{r}'|} \mathrm{d} \Gamma'  - \nonumber \\
&\sum_{j=1}^{N_p} \phi_1(\mathbf{r}_j) \int_{\Gamma_j} \frac{\partial}{\partial \mathbf{n}} \left[ \frac{e^{-\kappa |\mathbf{r}_i-\mathbf{r}'|}}{4\pi|\mathbf{r}_i - \mathbf{r}'|} \right] \mathrm{d} \Gamma'  = 0,
\end{align}

\noindent where $\mathbf{r}_i$ corresponds to the center of panel $\Gamma_i$, the collocation panel. 

After collocating on all panels, we obtain the following $2N \times 2N$ linear system of equations:

\begin{eqnarray} \label{eq:matrix}
 \left[
    \begin{matrix} 
       \frac{1}{2} I + K_L & -V_L \\
       \frac{1}{2} I - K_Y & \frac{\epsilon_1}{\epsilon_2}V_Y \\
    \end{matrix}
    \right] \left[ 
    \begin{matrix} 
       \phi_1 \\
       \frac{\partial \phi_1}{\partial \mathbf{n}} \\
    \end{matrix} 
     \right] =   
    \left[
    \begin{matrix} 
       Q \\
       0 \\
    \end{matrix}
    \right],
\end{eqnarray}

\noindent where,
\begin{align} \label{eq:matel}
 K_{L,ij} &= \int_{\Gamma_j} \frac{\partial}{\partial\mathbf{n}} \left( \frac{1}{|\mathbf{r}_{ij}|} \right) \mathrm{d} \Gamma  \nonumber &\quad
 V_{L,ij} &= \int_{\Gamma_j} \frac{1}{|\mathbf{r}_{ij}|} \mathrm{d} \Gamma \\ \nonumber
  K_{Y,ij} &= \int_{\Gamma_j} \frac{\partial}{\partial\mathbf{n}} \left( \frac{\exp (-\kappa |\mathbf{r}_{ij}|)}{|\mathbf{r}_{ij}|} \right) \mathrm{d} \Gamma \nonumber &\quad
 V_{Y,ij} &= \int_{\Gamma_j} \frac{\exp (-\kappa |\mathbf{r}_{ij}|)}{|\mathbf{r}_{ij}|}  \mathrm{d} \Gamma \\ 
 Q_i &= \sum_{k=0}^{N_c} \frac{q_k}{\epsilon_1} \frac{1}{|\mathbf{r}_{ik}|}.
 \end{align}

Here, $V$ and $K$ are the discretized versions of the single- and double-layer potentials, respectively, for the case of constant elements. The unknown vector in \eqref{eq:matrix} corresponds to $\phi_1$ and $\frac{\partial \phi}{\partial \mathbf{n}}$ evaluated at the center of each panel.
Once we solve the linear system, we can use the discretized form of  \eqref{eq:reac_integral} to find $\phi_{\text{reac}}$, the potential due to the solvent polarization, at the locations of $q_k$. We then compute the solvation energy from \eqref{eq:solv}.

\paragraph{Multi-surface problems} \label{sec:bem_multi}

\begin{figure}[t]
   \centering
   \includegraphics[width=0.7\columnwidth]{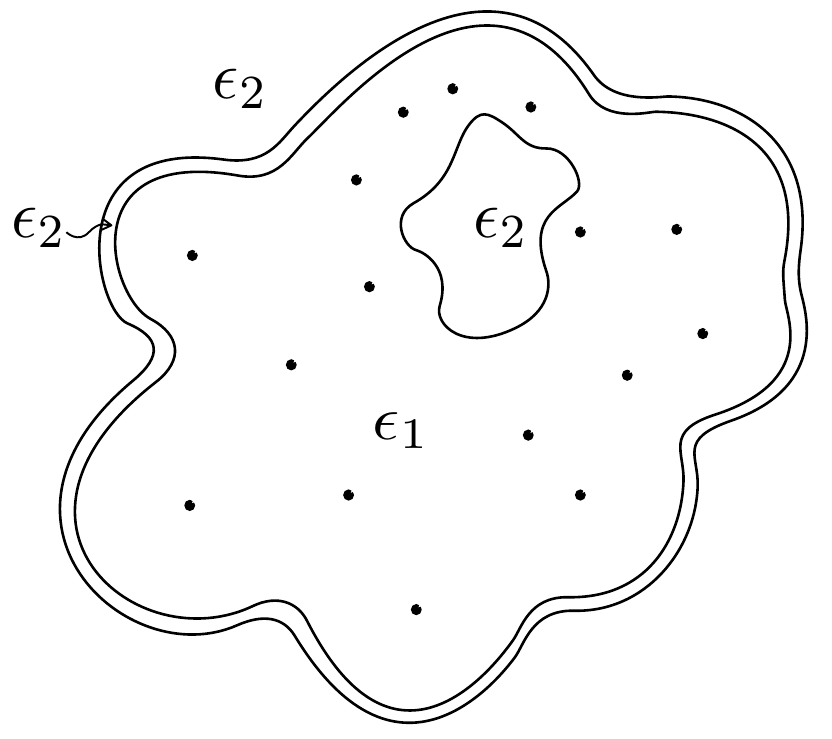} 
   \caption{Sketch of a molecule modeled with more than two regions: an ion-exclusion or Stern layer and one cavity are present.}
   \label{fig:molecule_multi}
\end{figure}

Following Altman et al.~\cite{AltmanBardhanWhiteTidor09}, \pygbe has the capacity to solve problems that involve multiple surfaces, such as when there are ion-exclusion (Stern) layers and solvent-filled cavities (depicted in Figure \ref{fig:molecule_multi}). 

An ion-exclusion or Stern layer is a 1 to 2~\AA-thick layer around the solvent-excluded surface that has the solvent's permittivity and no ions, and the potential is described by the Laplace equation.  The Stern layer compensates for the fact that the linearized Poisson-Boltzmann equation overestimates the concentration of ions near charged surfaces. This is an artifact of having a continuous representation of ions which does not consider steric effects, particularly important at charged interfaces. 

In the folded state, proteins may have cavities inside that are big enough to contain solvent molecules. Inside these cavities, the permittivity is that of the solvent, and the potential obeys the Laplace equation.  If, in addition, the cavity is large enough to hold ions, it can have its own interior Stern layer as well, and is governed by the linearized Poisson--Boltzmann equation.

Volumetric solvers include solvent-filled cavities and Stern layers automatically in the grid definition.  But in boundary-integral formulations, each region adds a surface that generates two extra equations, making the linear system of \eqref{eq:matrix} more complicated. In the design of \pygbe, we followed the method of Altman et al.~\cite{AltmanBardhanWhiteTidor09} to automatically determine the correct boundary-integral formulation for problems with multiple surfaces.   Readers are referred to that work for the details.

\subsection{Fast summation via treecode} \label{sec:treecode}

The treecode is an algorithm that reduces the complexity from $O(N^2)$ to $O(N \log N)$ for $N$-body calculations like evaluating the following sum in all points $i$:

\begin{equation} \label{eq:nbody}
V(\mathbf{x}_i) = \sum_{j=1}^{N} q_j \psi(\mathbf{x}_i, \mathbf{y}_j).
\end{equation}

\noindent It is based on clustering the ``sources'' $\mathbf{y}_j$ in boxes of an octree, and performing approximations when the ``target''  $\mathbf{x}_i$ is far away. The effect of the far-away sources is approximated using a series expansion about the cluster center. In this work, we use Taylor series for the expansions, writing for $\psi$:

\begin{equation} \label{eq:taylor}
 \psi(\mathbf{x}_i,\mathbf{y}_j) = \sum_{||\mathbf{k}||=0}^{P} \frac{1}{\mathbf{k}!} D_{\mathbf{y}}^{\mathbf{k}}(\mathbf{x}_i,\mathbf{y}_j) \psi(\mathbf{y}_j - \mathbf{y}_c)^\mathbf{k}.
\end{equation}

\noindent Here, $\mathbf{y}_c$ is the center of the box, $P$ the order of the expansion, $D_{\mathbf{y}}^{\mathbf{k}} = D_{y_1}^{k_1}D_{y_2}^{k_2}D_{y_3}^{k_3}$ the derivative operator, and $\mathbf{k} = (k_1, k_2, k_3)$, $\mathbf{k}! = k_1!k_2!k_3!$, $\mathbf{y}^{\mathbf{k}} = y_1^{k_1} y_2^{k_2} y_3^{k_3}$.

By rearranging terms and using \eqref{eq:taylor} in \eqref{eq:nbody}, we get the approximation of $V(\mathbf{x}_i)$ as

 \begin{equation} \label{eq:multexp}
 V(\mathbf{x}_i) = \sum_{||\mathbf{k}||=0}^{P} \frac{1}{\mathbf{k}!} \underbrace{D^{\mathbf{k}}_{\mathbf{y}} \psi(\mathbf{x}_i, \mathbf{y}_c)}_{a^{\mathbf{k}}} \underbrace{\sum_j q_j (\mathbf{y}_j - \mathbf{y}_c)^{\mathbf{k}}}_{m_c^\mathbf{k}},
 \end{equation}
 
 \noindent where $a^\mathbf{k}$ are the coefficients of the Taylor expansion and $m_c^\mathbf{k}$ the multipoles. Equation \eqref{eq:multexp} shows that the calculations involving the ``sources'' $\mathbf{y}_j$ can be decoupled from calculations involving the ``targets'' $\mathbf{x}_i$, which has the effect of reducing the complexity from $O(N^2)$ to $O(N \log N)$.
 
The octree is built stipulating that boxes at the lowest level ---called twigs or leaves--- should have less than \textsc{ncrit} sources. Once built, the tree is  traversed for each target starting from the root box (corresponding to the whole domain) querying if the box is far enough to use the approximation from  \eqref{eq:multexp}, or else, to look at the child boxes. This process is  recursively repeated, ending at the lowest level of the tree, where the interactions are computed directly. The criterion to decide if a box is far enough from a target particle to use the series is called the multipole acceptance criterion (\mac), given by
 \begin{equation} \label{eq:MAC}
 \theta > \frac{r_b}{r_{tb}},
\end{equation}

\noindent where $r_b$ is the size of the box and $r_{tb}$ the target-box distance. We use the approximation in \eqref{eq:multexp} when the criterion in \eqref{eq:MAC} is met, with the value of $\theta$ usually chosen as either $\frac{1}{2}$ or $\frac{2}{3}$.

The multipole terms $m_c^\mathbf{k}$ in \eqref{eq:multexp} are calculated first at the twigs of the tree, and then these values are shifted and added as necessary to obtain the multipole terms of all boxes in upper levels.
The next step is to traverse the tree and compute the rest of Equation \eqref{eq:multexp} where the \mac is met. For this step, we need to calculate the coefficients of the Taylor expansion $a^\mathbf{k}$, using recursive relations for the Green's function of the Poisson equation \cite{DuanKrasny2001} and linearized Poisson-Boltzmann equation \cite{LiJohnstonKrasny2009}. The final step is to directly compute interactions for boxes that are too close to targets in order to meet the \mac.
 
 
 \paragraph{Treecode-accelerated BEM}
 
 A BEM solver using an iterative method for the linear system of equations requires multiple dense matrix-vector multiplications, which can be accelerated to $O(N \log N)$ with a reecode. 
 Each element of the BEM matrix is an integral that can be approximated with Gauss quadrature, and we can formulate a matrix-vector product as a $N$-body problem like the one shown in \eqref{eq:nbody}, where quadrature points serve as sources, collocation points as targets and the function $\psi$ is the free-space Green's function of the Laplace or linearized Poisson-Boltzmann equations.
 
Integrating accurately near the singularity of the free-space Green's function poses a major challenge in BEM
\pygbe  computes singular integrals with a semi-analytical technique presented by Zhu et al.\ \cite{ZhuHuangSongWhite2001} and defines a near-singular region where it places more Gauss quadrature points. In practice, the near-singular region is always contained in an area where the \mac criterion \eqref{eq:MAC} is not met, so special integration using more Gauss points and the semi-analytical techniques are implemented in the direct-interaction step of the treecode. 

\pygbe uses a Krylov method for the solution of the linear system. The solver needs to perform one evaluation of the treecode per iteration, but uses the same source and target points. Because of this, we need to compute the tree structure and interaction list only once. 

\subsection{Richardson extrapolation} \label{sec:richardson}

Richardson extrapolation is a fundamental technique used in numerical computing to obtain an estimated exact solution from a series of computations using consecutive resolutions from coarser to finer. To be correctly applied, all of the computations used need to be converging to the exact value at a constant rate, meaning, they are in the asymptotic range. Under these conditions, an estimate of the exact solution can be written as:

\begin{equation} \label{eq:rich_extr}
f_{\text{exact}} \approx f_1 + \frac{f_1 - f_2}{r^{p} - 1},
\end{equation} 

\noindent where $f_1$  and $f_2$ are the fine-grid and coarse-grid solutions, respectively; $r$ is the mesh refinement ratio ($h_2/h_1$, where $h$ can be spacing, area or volume) and $p$ is the order of convergence.

From the calculations alone, we cannot ensure that the chosen grids are in the asymptotic range.  Before using Equation \eqref{eq:rich_extr}, then, one should check that $f_1$ and $f_2$ are in that range. We can extract the \emph{observed} order of convergence from three grid resolutions refined with constant ratio $r$, that is: $r = h_2/h_1 = h_3/h_2$. In that case:

\begin{equation} \label{eq:observed_order}
p = \frac{\log \left( \frac{f_3 - f_2}{f_2-f_1} \right)}{\log (r)}.
\end{equation}

If the result from \eqref{eq:observed_order} matches the \emph{expected} order of convergence of the method, it indicates that numerical computations for $f_1$, $f_2$ and $f_3$ are in the asymptotic range. For this reason, we need in general three calculations in the asymptotic convergence range to obtain an extrapolated value with Richardson extrapolation.

\subsection{Verification and Reproducibility}

We previously verified the \pygbe code using analytical solutions and published a technical report \cite{CooperBarba-share154331}.
The Barba research group has a consistent policy of making science codes freely available, in the interest of reproducibility. In line with this, we release the entire code that was used to produce the results of this study. \pygbe is made available under the MIT open-source license and we maintain a version-control repository.\footnote{\url{https://bitbucket.org/cdcooper/pygbe}} We also release all the files needed for running the numerical experiments reported in this paper: input, configuration, and post-processing. To support our open-science and computational reproducibility goals, in addition to open-source sharing of the code and the running scripts, several of the plots themselves are also available and usable under a \ccby license, as indicated in the respective captions.

\section{Results using \pygbe and comparison with \texttt{APBS}}
\label{sec:results}

\paragraph{Rationale for the experiments and setup} 

In the communities of biochemistry and biophysics, \apbs is a trusted software for biomolecular electrostatics, actively developed since the late 1990s \cite{BakerETal2001}. The \apbs code uses a finite-difference solution to the Poisson-Boltzmann equation on a Cartesian volumetric mesh, with more recent versions of the code providing a finite-element solution \cite{BakerETal2001b}. Our  objective was to test \pygbe using actual molecular geometries, comparing the results with those obtained using \apbs. The comparisons address two questions:  the accuracy that is obtained with these two different models and codes, and the conditions under which \pygbe can be competitive in terms of computational effort. 

We set up the \pygbe and \apbs runs using the parameters listed in Table \ref{table:parameters}, unless otherwise noted, with \apbs running on 6 cores of an Intel Xeon X5650 \cpu\ and \pygbe running on one \nvidia Tesla C2075 \gpu ~card.
To generate meshes, we used the free \msms software \cite{MSMS} and to determine charges and van der Waals radii, we used the free software \texttt{pdb2pqr} \cite{Dolinsky04}.

\begin{table}[h]
  \centering
  \fontfamily{ppl}\selectfont
    \begin{tabular}{ c c c c c c c}
	\toprule
	$\epsilon_1$ & $\epsilon_2$ & $\kappa$  & $K$ & $P$  & $\gmres_{\text{tol}}$ \\
	\midrule
	$4$ & $80$ & $0.125$ & $1$ & $2$ & $10^{-4}$ \\	
	\bottomrule
    \end{tabular}
    \caption{Parameters for setting up the runs with \pygbe and \apbs.} 
    \label{table:parameters}
\end{table}

In Table \ref{table:parameters}, $\epsilon_1$ and $\epsilon_2$ represent the permittivity of the protein and solvent regions, respectively, $\kappa$ is the inverse of the Debye length, $K$ is the number of Gauss points per triangular element, $P$ is the order of the Taylor expansion used for the treecode and $\gmres_{\text{tol}}$ is the tolerance of the Krylov iterative solver.

The tests used three molecular systems: (i) a lysozyme molecule, (iii) a trypsin-\textsc{bpti} complex; and (iii) a peptide-\textsc{rna} complex, all considered with and without cavities and/or Stern layer. See Table \ref{table:moleculeprep} for the protein-data-bank entries. In addition, we performed convergence tests using Richardson extrapolation with both \pygbe and \apbs, for which the test problem was a spherical molecule with an embedded charge. 

\begin{table}[h]
  \centering
    \begin{tabular}{ c c c }
	\toprule
	Case & Name & PDB code\\
	\midrule
	(i) & lysozyme & 1HEL \\	
	(ii) & trypsin-BPTI & 3BTK \\
	(iii) & peptide-RNA & \cite{Garcia-garciaDraper2003} \\
	\bottomrule
    \end{tabular}
    \caption{The three molecular systems used in the tests, and the corresponding entry in the Protein Data Bank.} 
    \label{table:moleculeprep}
\end{table}

\subsection{Convergence tests with both \pygbe and \texttt{APBS}}

\begin{figure*}
\centering
   \subfloat[Solvation energy convergence with mesh refinement.]{\includegraphics[width=0.48\linewidth]{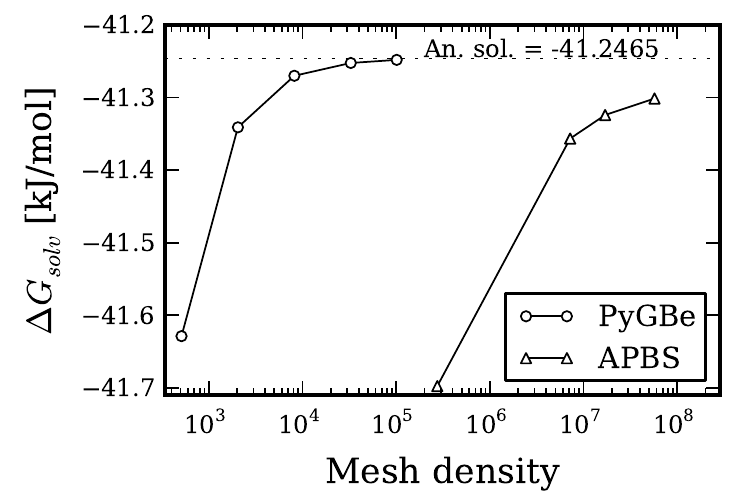} \label{fig:Esolv_sphere}}  
   \subfloat[Error convergence with mesh refinement.]{\includegraphics[width=0.48\linewidth]{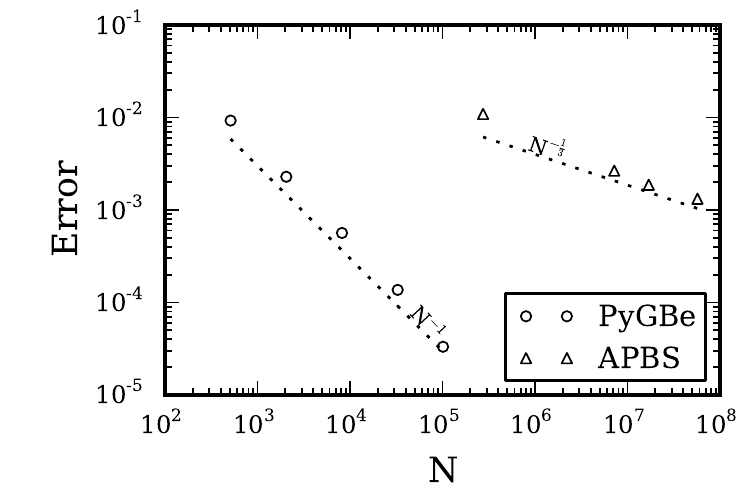} \label{fig:Error_sphere}}
   \caption{Convergence results for \pygbe and \apbs using a spherical molecule of radius 4\AA\ with a centered charge of 1e$^-$. The errors were calculated with respect to the corresponding extrapolated values (obtained from Richardson extrapolation). Data sets, figure files and plotting scripts available under \ccby \cite{CooperETal-share799692}.}
   \label{fig:sphere}
\end{figure*}

These results show the power of Richardson extrapolation. Using a spherical molecule of radius 4\AA\ with a centered charge and no ions in the solvent, we computed the electrostatic potential using \apbs and \pygbe with a variety of meshes and obtained the solvation energy. The parameters for these runs are detailed in Table \ref{table:sphere} and the results are shown in Figure \ref{fig:sphere}. The data sets, figure files and plotting scripts for Figure \ref{fig:sphere} are available under a \ccby license \cite{CooperETal-share799692}.

\begin{table}[h]
  \centering
  \fontfamily{ppl}\selectfont
    \begin{tabular}{ c c c c c c c}
	\toprule
	$\epsilon_1$ & $\epsilon_2$ & $\kappa$  & $K$ & $P$  & $\gmres_{\text{tol}}$ \\
	\midrule
	$4$ & $80$ & $0$ & $3$ & $12$ & $10^{-6}$ \\	
	\bottomrule
    \end{tabular}
    \caption{Parameters for the convergence tests using a spherical molecule.} 
    \label{table:sphere}
\end{table}

The plot in Figure \ref{fig:Esolv_sphere} corresponds to the solvation energy obtained with different meshes using the two codes. The analytical solution for our spherical test is $-41.2465$ [kJ/mol] and is represented by a dotted line on this plot.
The extrapolated values calculated using Equation \eqref{eq:rich_extr} are $-41.2471$ [kJ/mol] with \apbs and $-41.2467$ [kJ/mol] with \pygbe. These are both very close to the analytical solution and would require a very fine mesh to be obtained in a calculation with each code. 

Figure \ref{fig:Error_sphere} shows the error in solvation energy obtained with \pygbe and \apbs, calculated using the corresponding extrapolated value as the base solution. The plot  shows how error scales with the average area of the boundary elements for \pygbe and with mesh spacing for \apbs. 
The observed order of convergence calculated with Equation \eqref{eq:observed_order} was $1.014$ with respect to mesh spacing for \apbs and $1.013$ with respect to area for \pygbe. These values match the expected order of convergence for these methods, and thus we are confident that the computations are in the asymptotic convergence range. 

\subsection{Lysozyme } \label{sec:lys}

\begin{figure*}
   \centering
   \subfloat[Solvation energy convergence with mesh refinement.]{\includegraphics[width=0.48\linewidth]{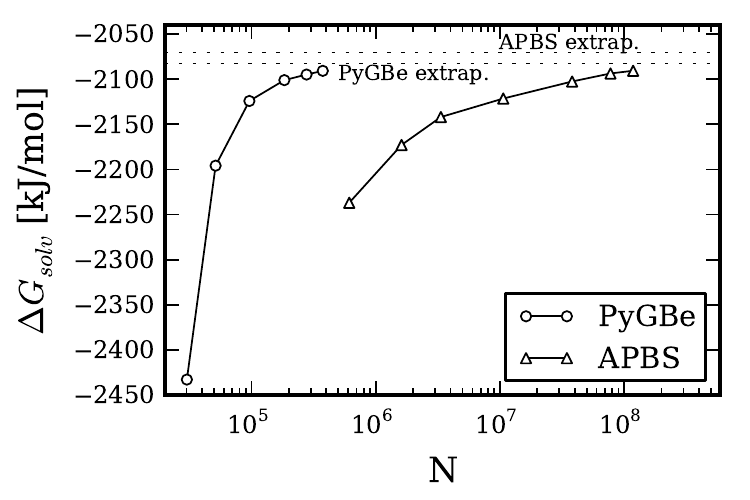} \label{fig:Esolv_lys}} 
   \subfloat[Error convergence with mesh refinement.]{\includegraphics[width=0.48\linewidth]{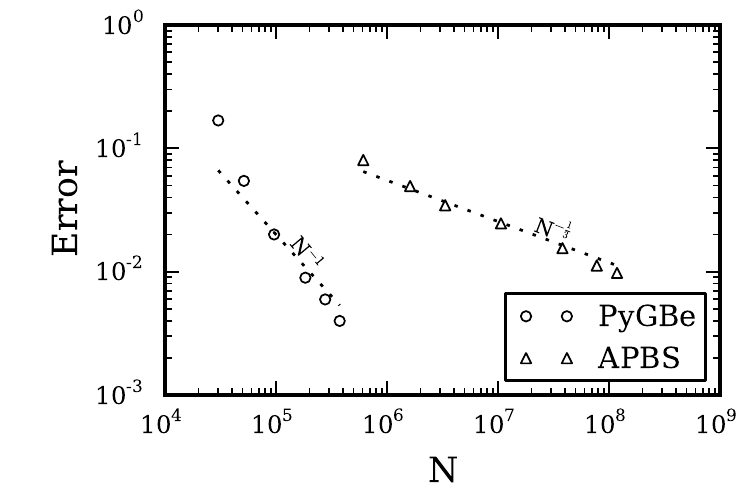} \label{fig:Error_lys}} \\
   \subfloat[Runtime vs.\ estimated error.]{\includegraphics[width=0.48\linewidth]{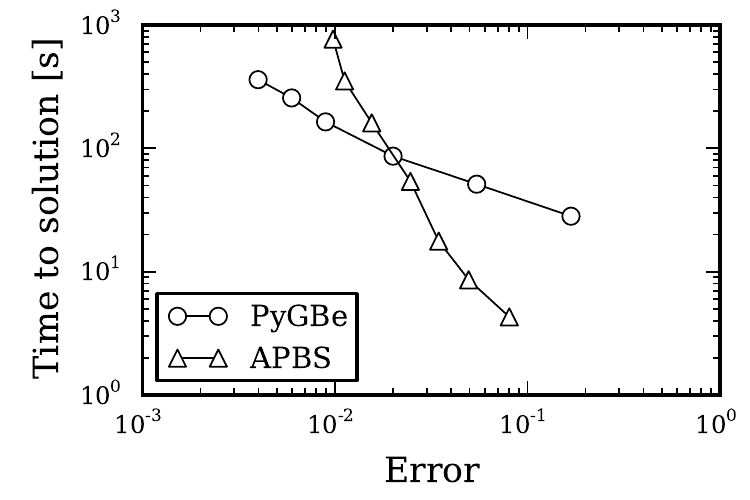} \label{fig:time_lys}} 
   \subfloat[Solvation energy comparison of various \pygbe models.]{\includegraphics[width=0.48\linewidth]{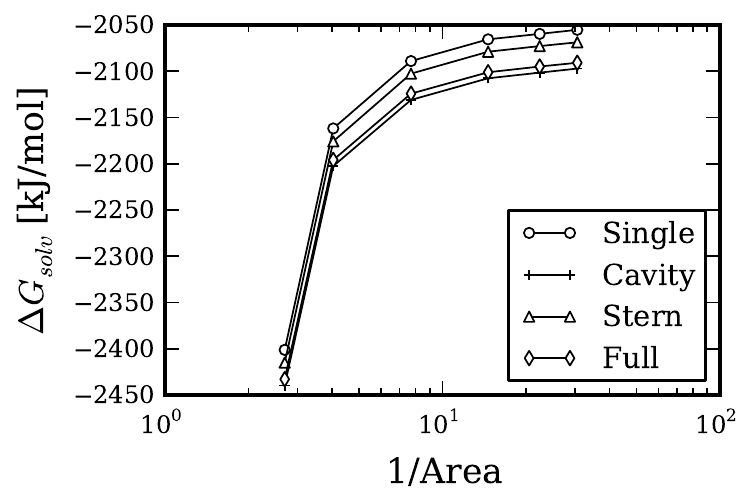} \label{fig:compare_lys}} \\ 
   \caption{Results for the lysozyme test, using \pygbe and \apbs. Data sets, figure files and plotting scripts available under \ccby \cite{CooperETal-share799702}.}
   \label{fig:lys}
\end{figure*}

Calculating solvation energy for a lysozyme molecule (code 1HEL) with different mesh refinements, we obtained the results shown in Figure \ref{fig:lys}. This molecule is modeled with 1323 point charges, has 3 cavities, and we also considered a 2\AA-thick Stern layer. 
The structure for lysozyme was the only one not prepared using \texttt{pdb2pqr}; instead, we used the data from a previous publication \cite{YokotaETal2011a}.
The calculation parameters are detailed in Table \ref{table:lys_param}, where: $K_{\text{fine}}$ is the number of Gauss points in the near-singular region, $N_\text{k}$ the number of Gauss points per side of the triangle for the semi-analytical integral technique used for the singular elements, \texttt{thres} the distance from the singularity defining the near-singular region ---with L being a characteristic length of the integration panel--- and $\theta$ the multipole-acceptance criterion of the treecode. The data sets, figure files and plotting scripts for Figure \ref{fig:lys} are available under a \ccby license \cite{CooperETal-share799702}.

\begin{table}[h]
  \centering
    \begin{tabular}{ c c c c}
	\toprule
	$K_\text{fine}$ & $N_\text{k}$ & $\texttt{thres}$  & $\theta$ \\
	\midrule
	$19$ & $5$ & $1.25L$ & $0.6$ \\	
	\bottomrule
    \end{tabular}
    \caption{Calculation parameters for the lysozyme case.} 
    \label{table:lys_param}
\end{table}

Figure \ref{fig:Esolv_lys} shows the computed solvation energy for different mesh sizes. As the mesh density increases, the discretization elements become smaller and the solution tends towards an asymptote, represented by the dotted line, which we calculated with Richardson extrapolation. This value is $-2070.47$ [kJ/mol] for \apbs and $-2082.5$ [kJ/mol] for \pygbe using the full model (with cavities and Stern layer). 

The errors plotted in Figure \ref{fig:Error_lys} were calculated using the corresponding extrapolated values as the base solution. The fact that the error is scaling with area for \pygbe and mesh spacing for \apbs proves that those calculations are in the regime with expected convergence, and that the error is dominated by the discretization. It is important to mention that mesh density for \apbs corresponds to number of cells per unit volume, whereas for \pygbe it is number of boundary elements per unit area, thus we cannot directly compare the discretizations, and we simply used these metrics for graphical purposes to have both cases in one plot.

We measured the time to solution and plotted the comparison between \apbs and \pygbe for different errors in Figure \ref{fig:time_lys}, using the errors from Figure \ref{fig:Error_lys}. For low-accuracy calculations, \apbs is faster than \pygbe; however, \apbs has worse scaling, and they cross-over near the $2\%$-error mark. This indicates that a boundary element approach is more suitable when there is a need for more accuracy.

The results shown in Figure \ref{fig:compare_lys} examine the importance of considering cavities and Stern layers. They correspond to a mesh-refinement study of solvation energy with Lysozyme modeled in four different ways: considering only the dielectric interface (``Single''), the dielectric interface plus cavities (``Cavity''), dielectric interface with Stern layer and no cavities (``Stern'') and including cavities and Stern layer (``Full''). For the most refined mesh, the difference between the ``Single'' result and the ``Full'' results is $35$ [kJ/mol]. The Richardson extrapolated value using the single-surface model is $-2047.16$ [kJ/mol], which is also around $35$ [kJ/mol] away from the extrapolated value for the "Full" simulation.

\subsection{Trypsin-\textsc{bpti}} \label{sec:bpti}

Using the method explained in Section \ref{sec:bind_energy}, we obtained the binding energy of the trypsin-\textsc{bpti} complex (code 3BTK) using both \pygbe and \apbs (\textsc{bpti} stands for Bovine Pancreatic Trypsin Inhibitor).
We first computed the solvation energy for each case and then added the Coulombic energy separately. The Coulombic energies for the trypsin-\textsc{bpti} complex, receptor and ligand are shown in Table \ref{table:coulomb}.

The \pygbe runs were done with two models: a multi-surface model, where the protein is modeled considering cavities and a 2\AA-thick Stern layer, and a single-surface model, where only the dielectric interface or SES is considered. The parameters used in these calculations are listed in Table \ref{table:bpti_param}. The data sets, figure files and plotting scripts for the results in this section are made available under a \ccby license \cite{CooperETal-share799703}.

\begin{table}[h]
  \centering
    \begin{tabular}{ c c c}
	\toprule
	\multicolumn{3}{c} {Coulombic Energy [kJ/mol]} \\
	Complex & Trypsin & BPTI \\
	\midrule
	$-79763.68$ & $-62956.98$ & $-17046.18$ \\	
	\bottomrule
    \end{tabular}
    \caption{Coulombic energies for the trypsin-\textsc{bpti} complex.} 
    \label{table:coulomb}
\end{table}

\begin{table}[h]
  \centering
    \begin{tabular}{ c c c c}
	\toprule
	$K_\text{fine}$ & $N_\text{k}$ & $\text{thres}$  & $\theta$ \\
	\midrule
	$19$ & $5$ & $2L$ & $0.5$ \\	
	\bottomrule
    \end{tabular}
    \caption{Calculation parameters for trypsin-\textsc{bpti} tests.} 
    \label{table:bpti_param}
\end{table}

\paragraph{Trypsin-\textsc{bpti} complex} \label{sec:complex}

\begin{figure}
   \centering
   \subfloat[Solvation energy convergence with mesh refinement.]{\includegraphics[width=0.98\columnwidth]{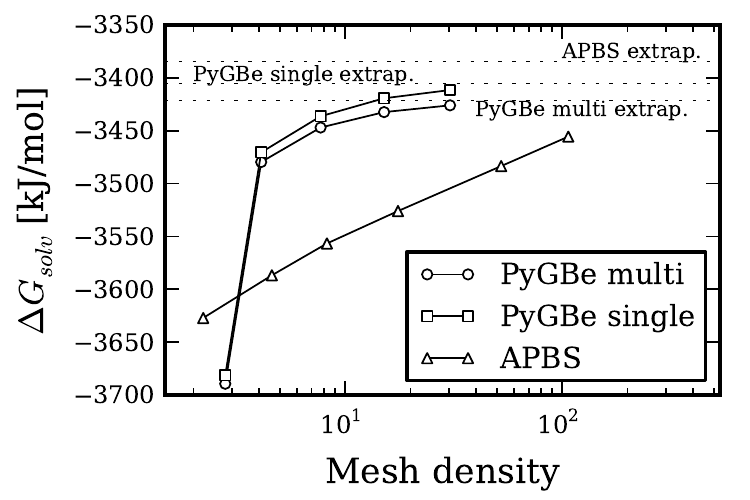} \label{fig:Esolv_bpti_com}} \\
   \subfloat[Error convergence with mesh refinement.]{\includegraphics[width=0.98\columnwidth]{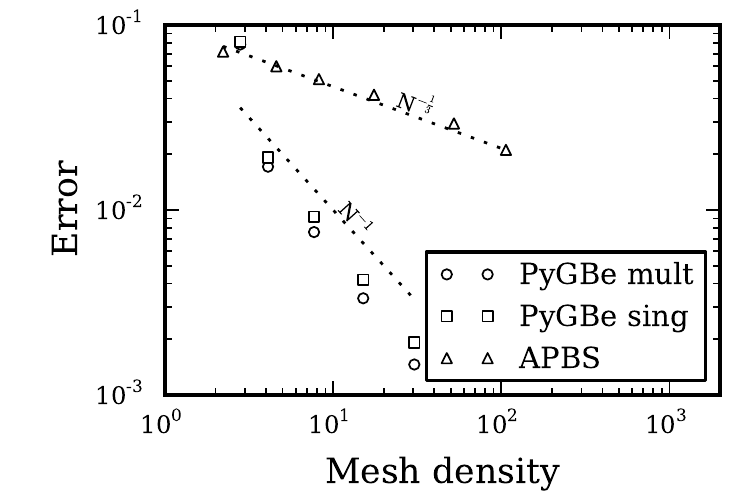} \label{fig:Error_bpti_com}} \\
   \subfloat[Runtime vs.\ estimated error.]{\includegraphics[width=0.98\columnwidth]{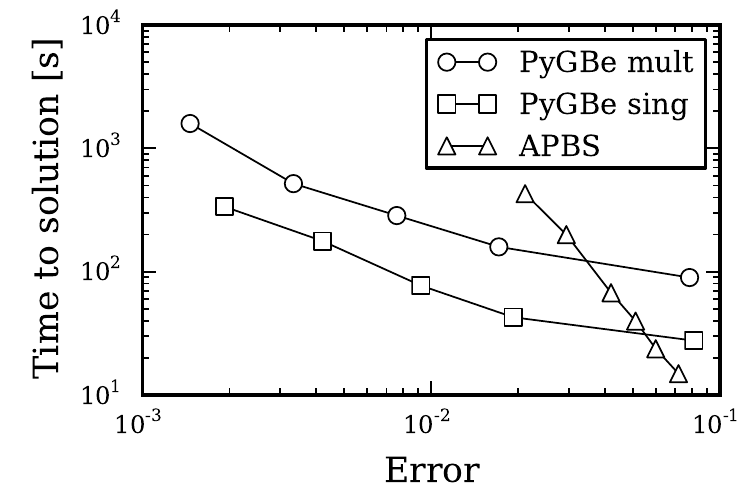} \label{fig:bpti_com_time}} \\
   \caption{\apbs and \pygbe results for trypsin-\textsc{bpti} complex. Data sets, figure files and plotting scripts available under \ccby \cite{CooperETal-share799703}.}
   \label{fig:bpti_complex}
\end{figure}

Figure \ref{fig:bpti_complex} shows the results of computing solvation energy for the trypsin-\textsc{bpti} complex, which we later use for obtaining binding energy. We modeled the trypsin-\textsc{bpti} complex using 4074 point charges with an Amber force field, and it contains 7 cavities.

The solvation-energy values obtained with different mesh densities are plotted in Figure \ref{fig:Esolv_bpti_com}. The dotted lines represent the extrapolated values obtained using Richardson extrapolation, and shown in Table \ref{table:bpti_extra} for each case: \apbs, multi-surface \pygbe and single-surface \pygbe. 
We calculated the errors shown in Figure \ref{fig:Error_bpti_com} using the extrapolated values in Table \ref{table:bpti_extra} as the base solution. The errors seem to be scaling with mesh spacing for \apbs and triangle area for \pygbe, at the expected rates.

Figure \ref{fig:bpti_com_time} shows a plot of time to solution with respect to estimated error. It displays the same behavior as the lysozyme molecule in Figure \ref{fig:time_lys}, where for low-accuracy calculations a volumetric approach is faster, but scales worse than the boundary integral approach, crossing over at a level of about 3\% error. Of course, the single-surface model requires less elements and takes less time to solve compared to the multi-surface model.

\paragraph{Trypsin} \label{sec:trypsin}

\begin{figure}
   \centering
   \subfloat[Solvation energy convergence with mesh refinement.]{\includegraphics[width=0.98\columnwidth]{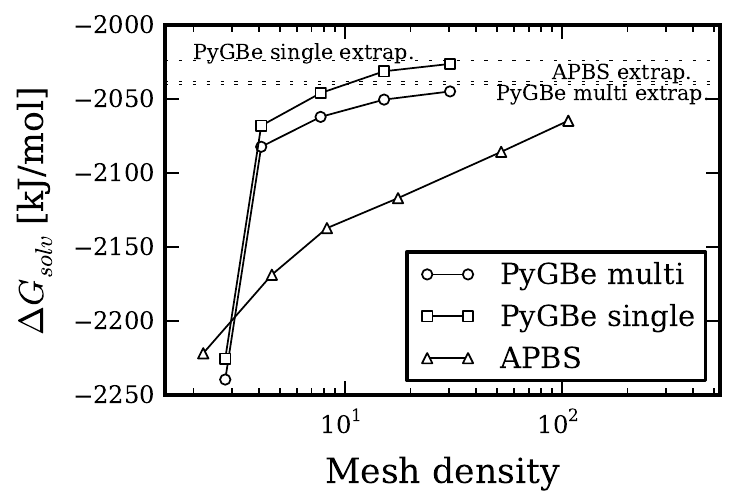} \label{fig:Esolv_bpti_rec}} \\
   \subfloat[Error convergence with mesh refinement.]{\includegraphics[width=0.98\columnwidth]{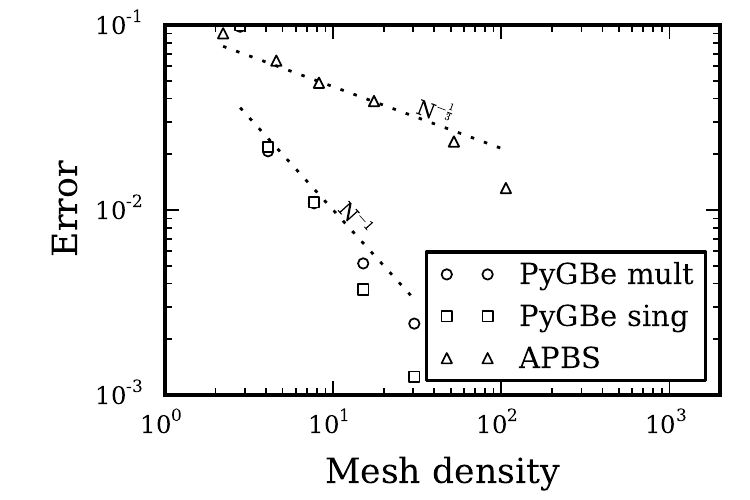} \label{fig:Error_bpti_rec}} \\
   \subfloat[Runtime vs.\ estimated error.]{\includegraphics[width=0.96\columnwidth]{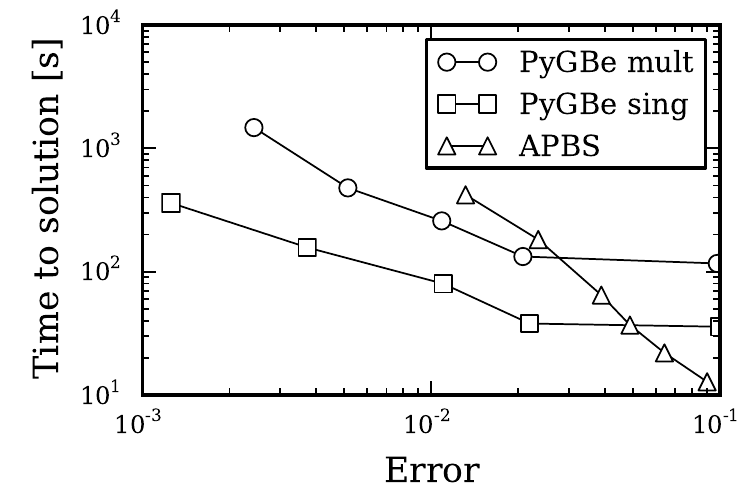} \label{fig:time_rec}} \\
   \caption{\apbs and \pygbe results for trypsin as receptor of trypsin-\textsc{bpti} complex.  Data sets, figure files and plotting scripts available under \ccby \cite{CooperETal-share799703}.}
   \label{fig:bpti_receptor}
\end{figure}

Figure \ref{fig:bpti_receptor} shows results of solvation energy for trypsin, which we later use for calculating binding energy. We model the trypsin using 3220 point charges with an Amber force field, and it contains 6 cavities.

Convergence of the solvation energy with mesh density is plotted in Figure \ref{fig:Esolv_bpti_rec}. The dotted lines represent the extrapolated values using Richardson extrapolation, listed in Table \ref{table:bpti_extra}.
We calculated the errors shown in Figure \ref{fig:Error_bpti_rec} using the extrapolated values in Table \ref{table:bpti_extra} as the base solution. The errors seem to be scaling with mesh spacing for \apbs and with area for \pygbe, at the expected rates.

The plot of time to solution versus error in Figure \ref{fig:time_rec} shows the same behavior as the lysozyme molecule in Figure \ref{fig:time_lys} and the trypsin-\textsc{bpti} complex in Figure \ref{fig:bpti_com_time}, where time to solution for the boundary element method has better scaling with error.

\paragraph{Bovine pancreatic trypsin inhibitor (\textsc{bpti})} \label{sec:bpti_sub}

\begin{figure}
   \centering
   \subfloat[Solvation energy convergence with mesh refinement.]{\includegraphics[width=0.98\columnwidth]{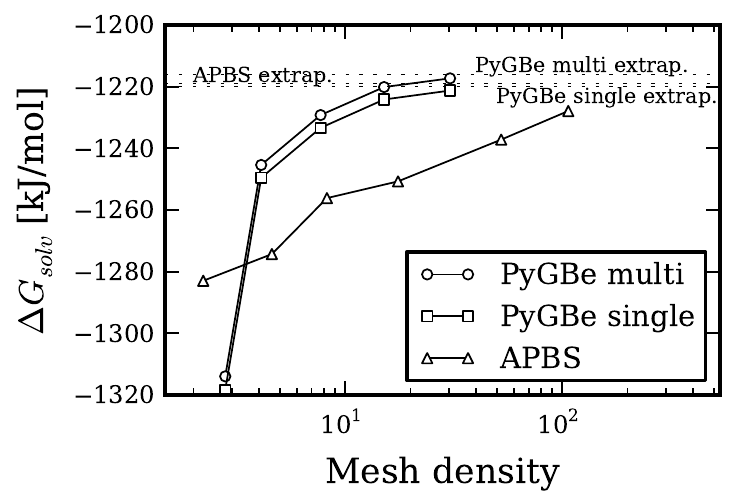} \label{fig:Esolv_bpti_lig}} \\
   \subfloat[Error convergence with mesh refinement.]{\includegraphics[width=0.98\columnwidth]{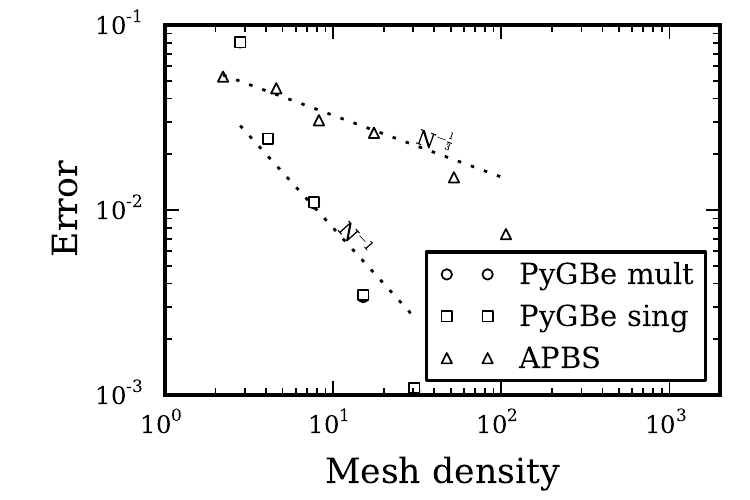} \label{fig:Error_bpti_lig}} \\
   \subfloat[Runtime vs.\ estimated error.]{\includegraphics[width=0.96\columnwidth]{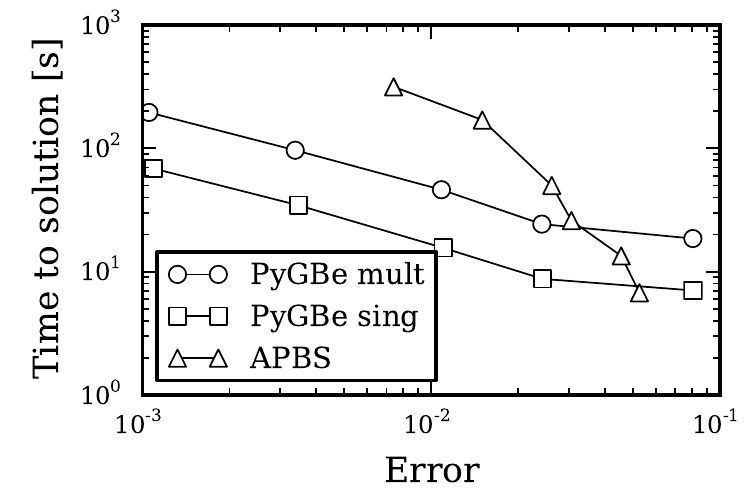} \label{fig:time_lig}} \\
   \caption{\apbs and \pygbe results for \textsc{bpti} as ligand of trypsin-\textsc{bpti} complex.  Data sets, figure files and plotting scripts available under \ccby \cite{CooperETal-share799703}.}
   \label{fig:bpti_ligand}
\end{figure}

Figure \ref{fig:bpti_ligand} shows results of solvation energy calculations for bovine pancreatic trypsin inhibitor (\textsc{bpti}), which is modeled using 854 point charges with a Amber force field, and no cavities.

The plot of solvation energy for different values of mesh density in on Figure \ref{fig:Esolv_bpti_lig}, where the dotted lines represent the extrapolated value using Richardson extrapolation, shown in Table \ref{table:bpti_extra}.
We calculated the errors shown in Figure \ref{fig:Error_bpti_lig} using the extrapolated values in Table \ref{table:bpti_extra} as the base solution. The errors seem to be scaling with mesh spacing for \apbs and area for \pygbe, in the expected rates.
Figure \ref{fig:time_lig} shows the plot of time to solution vs.\ error. As before, time to solution for the boundary element method has better scaling with error.

\paragraph{Trypsin-\textsc{bpti} binding energy} \label{sec:bind}

\begin{figure}[h]
   \centering
   \includegraphics[width=0.98\columnwidth]{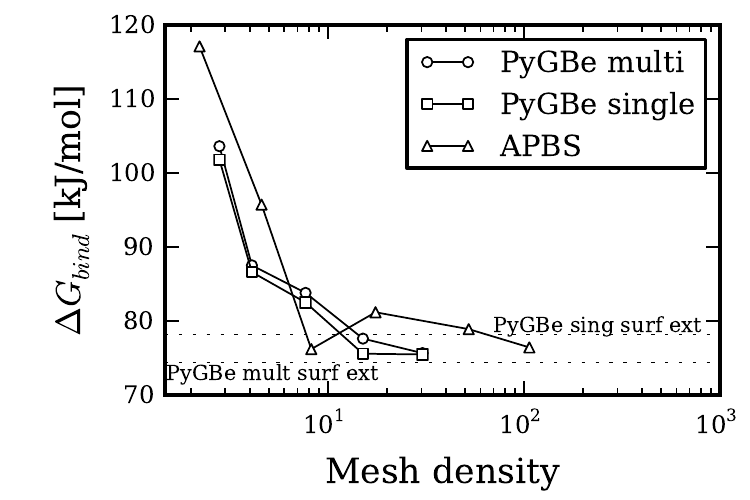} \label{fig:Ebind_bpti}
   \caption{\apbs and \pygbe results for binding energies of the trypsin-\textsc{bpti} complex. The multi-surface model of \pygbe considers cavities and Stern layer, and the single-surface model considers neither.  Data sets, figure files and plotting scripts available under \ccby \cite{CooperETal-share799703}.}
   \label{fig:bpti_bind}
\end{figure}

Binding energies can be computed from the solvation energies as detailed in Equations \eqref{eq:bind_energy} and \eqref{eq:total_energy}. Figure \ref{fig:bpti_bind} shows the computed  binding energy for different values of mesh density. Using the extrapolated values of solvation energy for \pygbe with multi- and single-surface models, shown in Table \ref{table:bpti_extra}, we calculated  binding energies. These were $74.42$ [kJ/mol] for the multi-surface model and $78.15$ [kJ/mol] for the single-surface model, and are represented in Figure \ref{fig:bpti_bind} by the dotted lines.

\begin{table}[h]
  \centering
    \begin{tabular}{ c c c c}
	\toprule
	& \multicolumn{3}{c} {Solvation energy [kJ/mol]} \\
	Code & Complex & Trypsin  & BPTI \\
	\midrule
	\apbs & $-3384.08$ & $-2037.88$ & $-1218.86$ \\
	\pygbe multi & $-3421.0$ & $-2039.92$ & $-1216.02$ \\	
	\pygbe single & $-3405.0$ & $-2023.74$ & $-1219.93$ \\	
	\bottomrule
    \end{tabular}
    \caption{Extrapolated solvation energies.} 
    \label{table:bpti_extra}
\end{table}

\subsection{Peptide-\textsc{rna} complex} \label{sec:pepRNA}

\begin{figure}[h]
   \centering
   \includegraphics[width=0.98\columnwidth]{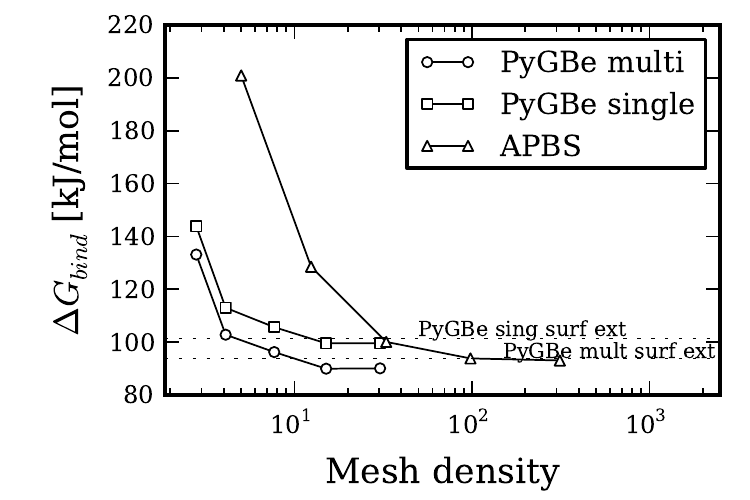}
   \caption{\apbs and \pygbe results for binding energies of a 22-residue long $\alpha$ helix peptide with ``box B'' hairpin structure of \textsc{rna}. \pygbe runs used a multi-surface model (considers Stern layer), and a single-surface model. Data sets, figure file and plotting script available under \ccby \cite{CooperETal-share799704}.}
   \label{fig:pepRNA_bind}
\end{figure}

We performed the same set of tests than in Section \ref{sec:bpti} but for a 22-residue long $\alpha$-helical  peptide of protein $\lambda$ bound the ``box B'' \textsc{rna} hairpin structure \cite{Garcia-garciaDraper2003}. 
In this case, the complex has 998 atoms, the \textsc{rna} 619 atoms and the peptide 379 atoms, and the structures are available online.\footnote{\url{http://www.poissonboltzmann.org/apbs/examples/}} Neither molecule has cavities and \pygbe runs used the multi-surface model with a 2\AA-thick Stern layer. Figure \ref{fig:pepRNA_bind} shows the binding energy of this complex for different mesh densities using \apbs and \pygbe with a multi-surface and single-surface model. The dotted lines correspond to binding energies calculated using the extrapolated values of solvation energy for the multi-surface model ($93.7$ [kJ/mol]) and single-surface model ($101.29$ [kJ/mol]). The data sets, figure files and plotting scripts for the results in this section are made available under a \ccby license \cite{CooperETal-share799704}.

\section{Discussion}
\label{sec:discussion}

The results presented in the previous section show that the extrapolated values for \pygbe and \apbs are converging to values that differ by $\sim 0.5\%$. Some difference is to be expected, because \apbs is a volumetric, Cartesian-mesh solver that constrains the molecular-surface definition to a staircase representation,  quite different from the surface-mesh representation used in \pygbe. Also, for volumetric approaches, point charges inside the molecule in general do not coincide with the mesh, and the required interpolation introduces smoothening of the point charges. 

Figure \ref{fig:compare_lys} shows the impact of including cavities and Stern layers in solvation-energy calculations, comparing results for lysozyme ---a molecule with 3 cavities--- modeled with and without these features. The extrapolated values of the ``Single" and ``Full" models differ by $35$[kJ/mol], which is a $\sim 2\%$ difference. This suggests that, for calculating solvation energy, if the required accuracy allows errors larger than $2\%$, then the simpler, single-surface model can be used, requiring less discretization elements and less runtimes. 

In the lysozyme test, we found also that the solvation energy is more affected by the presence of cavities than a Stern layer (Figure \ref{fig:compare_lys}). And as expected, results with \apbs  match the multi-surface \pygbe model better, since \apbs considers a Stern layer and cavities automatically. 

Although $35$ [kJ/mol] may be a small amount in the context of solvation energies, it may become important for derived quantities, for example, binding energies. Binding energy results from the difference between two solvation energies, which may be small. In the trypsin-\textsc{bpti} complex, the binding energy difference between multi- and single-surface models was only $4$ [kJ/mol], or $1.5RT$, with $R$ being the gas constant and $RT$ representing thermal fluctuations. This value is very close to thermal fluctuations, and corresponds to only $\sim5\%$ of the total binding energy, indicating that for this case a single-surface model is good enough. On the other hand, the peptide-\textsc{rna} complex of Section \ref{sec:pepRNA} shows a larger difference in binding energy when a Stern layer is considered (the components here have no cavities) of around $8$ [kJ/mol] or $3RT$, which corresponds to $\sim10\%$ of the total binding energy. Our results suggest that if the size difference between ligand and receptor is large, like in the trypsin-\textsc{bpti} complex, the resulting binding energy is less sensitive to Stern layers or cavities. However, for more comparable sizes in the molecules, like the peptide-\textsc{rna} complex, using all features becomes important. We think this happens because the receptor looks almost identical to the complex when it is much larger than the ligand, making the error introduced by using a single-surface model in the solvation energy calculation very similar for the receptor and complex, and being subtracted out in the binding energy calculation of Equation \eqref{eq:bind_energy}.

Our results in Section \ref{sec:results} also include time to solution. Timings for \pygbe do not include the mesh generation time with \msms, this time is negligible compared to the total time to solution. The trend of the time to solution plots is very similar in all cases, where lower-accuracy calculations are faster using a volumetric approach, but the scaling of the boundary integral technique with accuracy is better and they cross over at around $1\%$ error. This indicates that the choice of solver should be made considering the required accuracy of the application.

\section{Conclusion}
\label{sec:conclusion}

In this paper, we studied the use of a boundary element method to solve biomolecular systems in the context of implicit-solvent models. We noted that BEM, compared to volumetric-based methods, has difficulties to model situations where more than one surface is required to include all features of the biological system. The importance of including those features has not been treated before in the literature. We found that the answer is problem-dependent, and we noticed that, for example, in binding-energy calculations involving a ligand and receptor of comparable sizes, cavities and Stern layers should be included. Our work also included a comparison of the time-to-solution between \apbs and \pygbe. The results suggest that the choice of computational tool should be based on the desired accuracy, and when an error smaller than 1\% is desired, a boundary element approach is more adequate.

\bigskip

The results for this paper using \pygbe were obtained with version \texttt{e34c338} of the code, available from the version-controlled repository at \url{https://bitbucket.org/cdcooper/pygbe}. Input files, meshes and plotting scripts are also available \cite{CooperETal-share799692,CooperETal-share799702,CooperETal-share799703,CooperETal-share799704}.

\section*{Acknowledgements}
This work was supported by ONR via a grant from the Applied Computational Analysis Program, \# N00014-11-1-0356. LAB also acknowledges support from NSF CAREER award OCI-1149784 and from NVIDIA, Inc.\ via the CUDA Fellows Program.


\bibliographystyle{elsarticle-num}


\end{document}